\documentclass{aa}
\usepackage{graphicx}
\usepackage{txfonts}

\newcommand{\besancon}{Besan\c{c}on}

\begin{document}

\title{Probing the Canis Major stellar over-density as due to the Galactic warp}

\author{Y. Momany\inst{1}, 
S. R. Zaggia\inst{2}, 
P. Bonifacio\inst{2},
G. Piotto\inst{1}, 
F. De Angeli\inst{1}, 
L.~R.  Bedin\inst{1}, 
G. Carraro\inst{1} }
  
\institute {Dipartimento di Astronomia, Universit\`a di  Padova,   
Vicolo dell'Osservatorio 2, I-35122 Padova, Italy\\
\email{momany,piotto,carraro,deangeli,bedin@pd.astro.it }
\and
INAF - Osservatrio Astronomico di Trieste, via Tiepolo 11, 
I-34131 Trieste, Italy \\ 
\email{zaggia,bonifacio@ts.astro.it}
}

\date{Received April 21, 2004; accepted May 25, 2004}

\abstract{   Proper-motion,  star   counts  and   photometric  catalog
simulations are  used to explain the detected  stellar over-density in
the region  of Canis  Major, claimed  to be the  core of  a disrupted
dwarf galaxy (CMa, Martin  et al.\ \cite{martin04}, Bellazzini et al.\
\cite{luna04}), as due to the  Galactic warp and flare in the external
disk.
We  compare  the  kinematics  of  CMa  M-giant  selected  sample  with
surrounding Galactic disk stars in the UCAC2 catalog and find {\em no}
peculiar  proper   motion  signature:  CMa  stars   mimic  thick  disk
kinematics.
Moreover, when taking into account the  Galactic warp and flare of the
disk, 2MASS star count  profiles    {\em reproduce} the CMa    stellar
over-density.
This  star  count analysis  is  confirmed  by  direct comparison  with
synthetic  color-magnitude  diagrams  simulated with  the  \besancon\
models  (Robin et  al.\  \cite{robin03}) that  include  the warp  and
flare of the disk.
The presented evidence casts    doubt on the identification  of
the CMa over-density  as the core  of a disrupted Milky Way satellite.
This  however  does   not  make  clear the  origin   of over-densities
responsible  for the {\em ring}  structure in the anticenter direction
of the Galactic halo (Newberg et al.\  \cite{newberg02}; Yanny et al.\
2003; Zaggia et al.\ 2004, in preparation).

\keywords{Astrometry --- Galaxy: structure --- Galaxy: formation 
--- Galaxies: interactions} }

\authorrunning{Momany et al.}

\titlerunning{The CMa feature as due to the Galactic warp}

\maketitle

\section{Introduction}

Recent large scale surveys, in   the optical and near-infrared,   have
been excellent tools to constrain the structure and the star formation
history of the Milky Way and its satellite system. Growing evidence of
satellite accretion, stellar streams, and  sub-structures all point to
inhomogeneities  in both the  Galactic disk  and  halo.  The Sgr dwarf
spheroidal (Ibata et  al.\ \cite{ibata97}) is a  strong evidence of the
hierarchical formation in galaxies like the  Milky Way. Ever since its
discovery, the search  for extra-Galactic satellite remnants  has been
most appealing.
With the availability of the Sloan Digital Sky Survey, Newberg et al.\
(\cite{newberg02}) showed evidence   for at least 5 over-densities  in
the anticenter direction   of  the Galaxy.   Four of  these  features,
possibly part of a {\em ring}, were found close to the Galactic plane,
suggesting two possible origins: (1) the  remnant of a dwarf satellite
galaxy in the process of  disruption; or (2) a particular distribution
of   Galactic disk  stars.  The   ring  structure  has been  confirmed
kinematically  by Yanny et al.\  (\cite{yanny03}) and using 2MASS data
by Rocha-Pinto et al.\ (\cite{rocha03}).
Frinchaboy et al.\  (\cite{frin04}) and Crane et al.\ (\cite{crane03})
showed  the  existence  of  a structure     of stellar clusters   with
coordinated  radial velocities, further  confirming  the presence of a
possible  stellar  ring  at    a distance of  $\simeq18$~kpc    in the
anti-center direction.

Martin et al.\  (\cite{martin04}, M04), investigating the ring structure
with the 2MASS all-sky catalog, pointed to an elliptical-shaped stellar
over-density centered  at $l=240\degr$, $b=-7\degr$.  Accounting
for simulations of (a) past mergers thick disk formation (Abadi et al.\
\cite{abadi03}), and  (b) dwarf galaxies in-plane  accretion (Helmi et
al.\  \cite{helmi03}), M04  proposed this ``feature'' as  the core of a
satellite galaxy currently undergoing   an in-plane accretion,  namely
the \object{Canis Major dwarf spheroidal galaxy} (CMa).
In   a  companion   paper,   the  same   group   (Bellazzini  et   al.\
\cite{luna04})   searched  for   photometric  signatures  of   CMa  in
color-magnitude diagrams (CMDs), and best identified the red clump and
red  giant  branch of CMa in  the  CMDs of the  Galactic open clusters
\object{NGC 2477}   and Tom~1.    Forbes,  Strader \& Brodie (\cite{forbes04})
studying the age--metallicity  relation   for  CMa  probable   cluster
members pointed out  a clearly  distinct  extragalactic origin of  CMa
debris.  On  the other hand, Kinman,  Saha \&  Pier (\cite{kinman04})
did not  find  any over-density of   RR Lyrae stars in an  anti-center
field  of   the Galaxy.    This   contrasts with  the  Zinn    et al.\
(\cite{zinn03})   study  who  possibly   identified  CMa  stars in the
so-called southern arc of the CMa ring.

Accreting Galactic satellites could be responsible for the creation of
the  observed    Milky  Way warp  (Castro-Rodr{\'{\i}}guez    et   al.
\cite{castro02}).  Evidence for  the existence of  a Galactic warp for
the  inter-stellar gas   and dust  are    long dated  since  early  HI
observations (Oort, Kerr \& Westerhout \cite{oort58}), and more recent
inter-stellar diffuse dust emission (Freudenreich et al.\
\cite{freud94} and Freudenreich \cite{freud96}) studies.  The warp has
been  also detected  in  stellar  star counts and    can be seen  as a
systematic  variation  of  the   mean  disk latitude    with longitude
(Djorgovski \&  Sosin \cite{djor89}): observationally,   the warp is an
upward  bending from   the  Galactic  plane in the    first and second
galactic  longitude quadrants  $(0\degr\le l  \le180\degr)$, and
downward  in    the third  and  fourth  ones   ($180\degr\le l  \le
360\degr$).  {\em The warp  feature  is therefore important in  the
direction of CMa and  has to be accounted  for when asymmetries across
the Galactic plane are investigated}.

A comprehensive formation  scenario for the features found  in CMa and
their connection with other galactic structures seriously call for the
determination  of  basic  observational  properties of  CMa  such  as:
metallicity, proper motion (pm),  radial velocity etc.  In the present
paper  we address the  mean pm  of stars  in the  direction of  CMa in
search for  any kinematic signature.   We also investigate  a possible
connection of the CMa over-density with the {\em warp} and {\em flare}
of the  Galaxy (Robin et  al.\ \cite{robin03} and  references therein)
which manifest their maximum extent in the CMa region.

\section{CMa proper-motion}

In the  absence of specific pm studies  of CMa, in this  paper we make
use of the  recent USNO CCD Astrograph Catalog  second release (UCAC2,
Zacharias  et  al.\  \cite{zach04})   to  investigate  a  possible  pm
signature  of the  CMa.  The  first epoch  data for  the pm  come from
different  catalogs including  Hipparcos/Tycho, AC2002.2,  as  well as
re-measured AGK2, NPM  and SPM plates.  In general,  Zacharias et al.\
estimate pm errors to be $1\div3$ mas/yr for 12$^{th}$ magnitude stars
and  about $4\div7$  mas/yr around  the 16$^{th}$  magnitude.   The pm
reference  is  absolute  and  it  has been  evaluated  using  external
galaxies.  A useful  feature of the UCAC2 catalog  is the inclusion of
the $J$, $H$ and $K$ magnitudes from the 2MASS survey.
We also  made use of the  on-line Milky Way  photometric and kinematic
simulations  based  on the  \besancon\  Galaxy  model  (Robin et  al.\
\cite{robin03}): its  main advantage is  the reduction of a  number of
free parameters  (e.g. the thin  disk scale height at  different ages)
and,  most interestingly,  the  inclusion of  {\em  warping} and  {\em
flaring} of the Galactic disk.

\begin{figure}
\centering \includegraphics[width=0.45\textwidth]{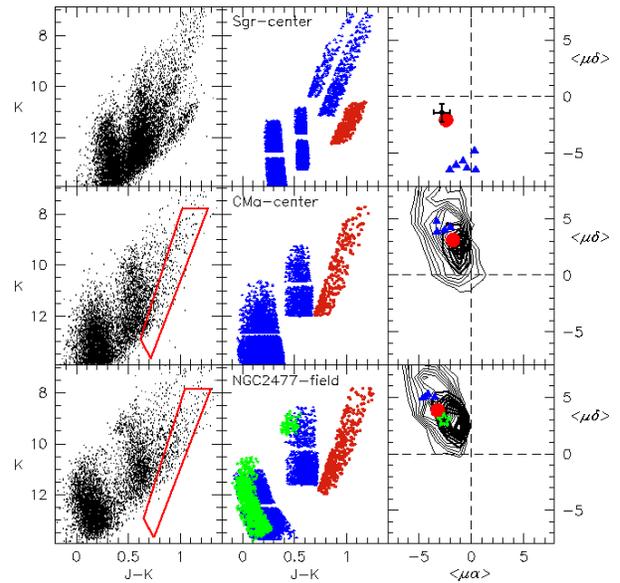}
\caption{2MASS  $K,(J-K)$  dereddend   CMDs  and  corresponding  UCAC2
proper-motion plots of the Sgr dwarf spheroidal, CMa and \object{NGC 2477}.  For
each row, from left to right: (1) a 1$\degr$ $K$, ($J-K$) 2MASS CMD
limited to stars with UCAC2 pm data based on the AC2002.2 catalog; (2)
the same  CMD but showing   only the selected  stellar populations for
which  a mean pm  is to  be  determined; and (3) the   mean pm of  the
selected   populations   following a  $2\sigma$-clipping  to eliminate
outliers (see online colored version of the Figure). }
\label{f_fig1}
\end{figure}

To estimate the pm of CMa  we extract UCAC2 $\mu \alpha$, $\mu \delta$
(in  this paper we  assume $\mu\alpha = \mu\alpha\,  cos\delta$), $J$,
and $K$  data in 3 fields  having a radius of  $1\degr$.  The first
centered on the  Sgr to be used for testing  the procedure, the second
on  CMa  and the  last  on the  open  cluster  \object{NGC 2477}.  Adopting  the
selection criteria outlined in Bellazzini  et al.\ and M04, we use the
$K$, ($J-K$)  diagrams to extract M-giants  {\it presumably} belonging
to CMa and estimate the mean UCAC2 pm.
Although  simple, this  approach  is  rather tricky.   As  it is  well
depicted by the study of Platais et al.\ (\cite{plat03}), the presence
of  residual  color/magnitude  terms  in  any pm  analysis  is  almost
unavoidable.  This  is specially true  when one deals with  an all-sky
catalog,  where the  {\em pm  is  derived from  different first  epoch
sources}.   Aware of  these facts,  the  minimal approach  was to  (i)
assess the reliability of the  UCAC2 measurements by a comparison with
fields of known pm, and (ii) limit the UCAC2 pm data to {\em only one}
first  epoch  catalog  in  order to  limit  internal  inhomogeneities.
Several  tests were  performed re-measuring  the mean  pm  of globular
clusters and  comparing these with  the values reported in  Dinescu et
al.\ (\cite{dine99}).   The overall agreement  has been very  good for
all measurable cases  ($16$ clusters with distances ranging  from 3 to
13  kpc),  with  an  overall  scatter  of  less  than  $2.5$~mas.   In
conclusion,  the best  confidence level  was found  with  the AC2002.2
dataset (see  Zacharias et al.\  2004): we restricted our  analysis on
it.

Fig.~\ref{f_fig1} presents CMDs and pm diagrams for 3 selected fields,
from   top    to   bottom:   {\em    Sgr-center},   {\em   CMa-center}
($l=240\degr$,   $b=-7\degr$)  and  {\em   NGC2477-field}.   The
stellar  populations selected in  the middle  panels are  mainly field
stars chosen to span all  possible color and magnitude ranges in order
to  check  for  any  pm  gradient.   The bulk  of  the  selected  disk
populations  with  $(J-K)<0.8$ are  expected  to  have a  heliocentric
distance of $<2$~kpc (open  triangles), while objects with $(J-K)>0.8$
and  $8\le K  \le14$ are  found mainly  at distances  $>2$~kpc (filled
circles).
On  the top  row of  Fig.~\ref{f_fig1} the  identification of  the Sgr
M-giants (filled  circles) is  straightforward, showing the  red giant
tip at $K\simeq11.0$.  The right  upper panel plots the estimated mean
pm  of the  6 presumably  {\it field}  populations  (filled triangles)
which           show           a           clustering           around
$<\mu\alpha,\mu\delta>\simeq(-1.0,-6.0)$~mas.   The dispersion  of the
mean pm of the field stars is due to the different nature and distance
of  the selected  objects.   The  dispersion of  the  6 field  samples
($\simeq2.0$~mas) can be used as  a conservative error estimate of our
mean pm determinations.   Relatively speaking, the mean pm  of the Sgr
red    giants   clearly    stands   out    at   $<\mu\alpha,\mu\delta>
\simeq(-2.5,-2.0)$~mas, and  is well separated  from surrounding field
stars.  This is due to the almost polar orbit of Sgr.
A filled square with error bars shows the mean value of the Sgr proper
motion as  derived by  Irwin et al.\  (\cite{irwin96}) from  HST data:
$<\mu\alpha,\mu\delta>=(-2.80\pm0.80,-1.40\pm0.80)$~mas.            The
excellent agreement between the  two values strengthens our confidence
in measuring pm  with UCAC2 up to a distance  of $\sim25$~kpc from the
Sun.

Turning our  attention to CMa,  we extract M-giants within  an oblique
stripe, as  done in Bellazzini  et (\cite{luna04}) in  the \object{NGC
2477} field.
CMa giants are plotted as filled  circles in the central panels of the
middle and  lower rows of  Fig.~\ref{f_fig1}.  Stars belonging  to the
open cluster were selected within  a radius of $0\fdg2$ (light starred
symbols), clearly showing the main sequence and red clump.
As in the  Sgr field, the mean pm of  disk stellar populations (filled
triangles)  in the  CMa-center  and \object{NGC  2477}  fields show  a
dispersion of $\pm2.0$~mas, with  a rather lower dispersion around the
\object{NGC 2477} as expected for its low latitude.
We   estimate    a   mean   pm   of   CMa-center    M-giants   to   be
$<\mu\alpha,\mu\delta> \simeq (-1.7,3.1)$~mas.  This is in the same pm
direction of disk stars, although slightly offsetted ($\sim1.5$~mas in
both $\mu\alpha$ and $\mu\delta$). 
To better  understand this  offset, on both  right panels  we overplot
contour  levels  of simulated  kinematic  stellar  catalogs using  the
\besancon\ Galactic  model.  The simulated catalogs  were divided into
two stellar  subgroups (as done  for the 2MASS $J,K$  color selection)
according  to their  heliocentric  distance: thin  line contours  show
stars within  2~kpc, i.e. $(J-K)<0.8$, while thick  line contours show
stars with  distance $>2$~kpc, i.e. $(J-K)>0.8$.  For each population,
the counts have been divided in 9 linear contour levels.
We  note that  the  above stated  value  of CMa-center  pm is  in fair
agreement  with expected  mean pm value   of the latter  group of disk
objects,  with  a   mean  distance  of  $\simeq   8$~kpc.   Thus, when
accounting for  a mean error of $\pm2.0$~mas  and an offset merely due
to distance projections,   the mean pm  of  CMa-center seems perfectly
compatible with outer disk dynamics in circular prograde motion.
This finding is in agreement with  the Crane et al.\ (2003) solution for
the kinematics  of the "Monoceros" structure  which appears compatible
with a disk component rotating at 220 km/s at a distance of 18~kpc and
showing a thick disk velocity dispersion.
Radial  velocity measurements and  a detailed  computation of  the CMa
orbit will  be presented in a  separate paper (Zaggia et  al., {\it in
preparation}).

\begin{figure}
\centering
\includegraphics[width=7.7cm]{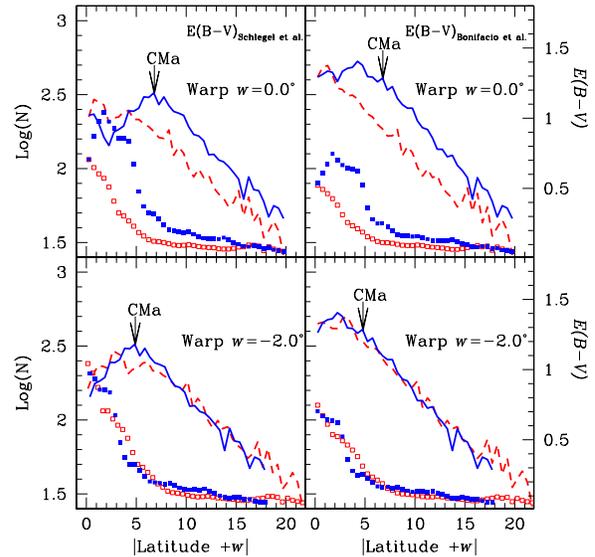}
\caption{Latitude profiles   of a 2MASS  M-giant  star sample  selected
around  CMa.  Upper-left  panel assumes a  North/South symmetry around
$b=0\degr$,    lower-left  panel assumes     a  warp  amplitude  of
$2\fdg0$ in the southern  direction.  Star counts in  left panels
are corrected for reddening using the Schlegel et al. maps. Right panels
show the same plots, except for correcting  the Schlegel et al. values
with the formula given in Bonifacio et al.}
\label{f_stripe}
\end{figure}

\section{CMa and the Galactic disk warp}

L{\'  o}pez-Corredoira  et al.\  (\cite{lopez02}) recently studied the
existence  of the warp in the  Galactic stellar populations using star
counts of old-population red clump giants selected from 2MASS CMDs.
Their Fig.~15, on  the  ratio of North and   South Galactic caps  star
counts  shows a clear sinusoidal  behavior  with longitude.  Excess of
star counts in the  northern hemisphere latitudes  was found for $l\le
180\degr$, and  an   opposite trend  for $l\ge180\degr$.    They
basically  concluded their analysis   {\em confirming the existence of
the warp in  the old stellar  population whose amplitude is coincident
with that of warped gas and young disk stars}.  They also identified a
strong flare, i.e.  a change in the scale-height, of the outer disk.
Previously,   Freudenreich  et al.\  (\cite{freud94})  using COBE/DIRBE
near-infrared data  found that the absolute   maximum amplitude of the
latitude displacement is $1\fdg5 \div 2\fdg0$.  
Interestingly, the over-density  in CMa and  {\em ``structure  A''} in
the symmetric quadrant ($0\degr<l<180\degr$)  (see Figs.~4 and  5
in  M04) coincide respectively with  the southern and northern regions
where the warp amplitude is strongest.
This clearly indicates that the warp and flare of the disk can heavily
affect star counts analysis of low latitude, distant regions.
To test the impact of Galactic warp  and flare near the core of CMa we
revisited  the analysis  performed by  M04.  A  suitable  CMa M-giants
sample was  extracted from  the 2MASS catalog  using the  same oblique
selection region depicted  in the CMDs of Fig.~1,  i.e.  {\em sampling
the region with CMa giants  excess}.  The sample was spatially limited
along a latitude stripe  between $-20\degr\le b \le 20\degr$, in
the   longitude  range  $235\degr\le   l  \le   245\degr$,  i.e.
corresponding to the CMa spatial FWHM, as derived in M04.
$J$ and $K$  magnitudes were corrected for reddening  in two different
ways using: (1) {\it raw} Schlegel et al.\ (\cite{schl98}) values, and
(2) {\it modified} Schlegel et  al.\ values as suggested in Bonifacio,
Monai  \&  Beers\  (\cite{boni00}).   Star  counts  were  computed  in
latitude  bins of $0\fdg5$  (the results  don't change  using the
$4\degr\times2\degr$ binning of M04).

In  Fig.~\ref{f_stripe} the upper-left panel    shows the folded  star
count profiles for  the  northern (dashed  line) and  southern  (solid
line) galactic caps.  In this  panel the assumed North/South  symmetry
is at $b=0\degr$ without   introducing any warping in the   Galaxy,
i.e.  with a warping angle $w=0.0$.   Also plotted are the Schlegel et
al.\ (1998) reddening mean values for the southern (filled rectangles)
and northern (open  rectangles)   hemispheres along the  same  spatial
region.  The CMa location is identified with a vertical arrow.
A  glance at Fig.~\ref{f_stripe}  shows that the  southern part of the
profile  is clearly   offsetted with  respect  to  the  northern part.
In  the lower-left panel of Fig.~\ref{f_stripe}  the  same star counts
are presented  assuming  a warp amplitude  of  $w=2\fdg0$  in the
southern direction,  i.e.  a North/South symmetry at $b=-2\fdg0$.
The curves plotted   in the  lower panel   {\em  show no}  significant
changes varying $w$ by $\pm0\fdg3$ around $w=-2\fdg0$.
The clearest  feature in  this  panel is  the almost  perfect symmetry
around $b=-2\fdg0$.  Moreover, the  North/South symmetry is  also
reflected in the two reddening curves which show a far better symmetry
down to the mid plane. As a result  of taking into account the disk
warp, one  notes that  although  the CMa over-density persists,  it is
{\em seriously weakened}.

Various authors have noted that the Schlegel et al.\ maps overestimate
reddening  for   $E(B-V)>0.2$,  in    particular,   Bonifacio et  al.\
(\cite{boni00}) proposed a linear correction [their formula (1)] which
lowers asintotically $E(B-V)$ of $\simeq35$\%.  The correction becomes
quite important  at  low latitude  with   a difference of   as much as
$\simeq0.2$~mag in $(J-K)$ for an $E(B-V)=1.0$.  The previous analysis
is repeated applying the Bonifacio et al.\ correction (right panels in
Fig.~\ref{f_stripe}).
Assuming a $w=-2\fdg0$ and correcting  the reddening values it is
quite evident that {\em both}  star counts and reddening profiles show
an excellent symmetry.  This   time however, the CMa  over-density has
almost disappeared.
The assumed  value of $w$  is in  good  agreement with that  found  by
Freudenreich et  al.\ (\cite{freud94},  their Fig.3)  and Freudenreich
(\cite{freud96}) for this particular zone.  These studies suggest that
the layers of  dust and neutral hydrogen  are similarly displaced from
the  Galactic plane.  As  regarding  to  stellar warp,  evidence of  a
similar  $w$ value has  been suggested in   the 2MASS analysis of L{\'
o}pez-Corredoira et al.\ (\cite{lopez02}).
Indeed,  for the  line of  sight at  $l=240\degr$  and galactocentric
distances in the  range $10\div14$~kpc they predict the  height of the
mid  plane (due  to the  warp)  to range  between $300-450$~pc.   This
translates into a warping angle of $w=1\fdg8 \div 2\fdg4$ in the South
direction.
In conclusion,  {\em  an appropriate  consideration  for the  warp and
flare completely erases any over-density in CMa}.

\begin{figure}
\centering
\includegraphics[width=8.0cm,height=8.0cm]{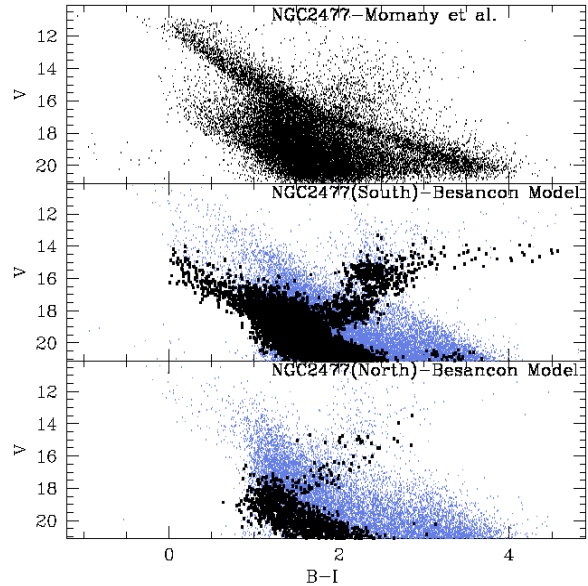}
\caption{$V$, ($B-I$) diagrams of NGC2477. Upper panel is from Momany
et al.\ (\cite{moma01}), middle panel is a \besancon\ simulated CMD in
an  equivalent       area   along    the  same     line      of  sight
($b=-5\fdg8$). Lower panel is  a   \besancon\ simulated CMD    at
positive latitude ($b=+5\fdg8$).}
\label{f_2477}
\end{figure}

Further support   of the   importance  of  the  warp  comes  from  the
comparison of observed and synthetic CMDs.
In Fig.~\ref{f_2477} we  compare a $V$, ($B-I$)  CMD of \object{NGC 2477} (upper
panel, Momany  et  al.\  \cite{moma01}) with  a  \besancon\  synthetic
catalog  (middle panel). There  is  an excellent agreement between all
observed features and simulated ones, except of course for the absence
of  the open cluster component   and differential reddening effects in
the  simulated CMD. Clearly, sequences   previously attributed to  CMa
(blue plume, young main sequence, red giants  and red clump) are fully
reproduced by the Galactic warped model.
Plotted   as  heavy symbols are   stars   with a heliocentric distance
between $7\div9$~kpc, a  mean metallicity of [Fe/H]$=-0.45\pm0.25$ and
a mean age  of  $5.0\pm1.5$~Gyr.  Curiously, similar parameters   have
been  derived   in  Bellazzini et al.\  (2004)   for  the CMa  stellar
populations: a mean heliocentric distance of  $8.3\pm 1.2$ kpc, a mean
metallicity in the range $-0.7\le$[M/H]$\le0.0$ and a mean age between
$2\sim7$  Gyr.  
On  the other  hand, the  Galactic warped  model shows  a considerable
decrease  (a  $\sim$10  factor)   of  the  stellar  component  between
$7\div9$~kpc in the NGC2477-North field (Fig.3, lower panel), i.e.  it
reproduces  quite well the  CMa South/North  over-density.  Similarly,
\besancon\  simulated  diagrams   reproduce  CMDs  of  the  background
populations  of  the  open   clusters  having  $-18\degr\le  b  \le
-14\degr$ (see  Sect.  4.3  in Bellazzini et  al. 2003),  showing a
substantial decrease of the warp background contribution.

{\em   In  conclusion}, accounting for the   Galactic  warp  and flare
explains   the  detection of   both the CMa  over-density  and related
stellar populations.  As a consequence, the CMa feature can not be the
progenitor of the {\it ring} found by Newberg et al.

\begin{acknowledgements}
We thank the anonymous referee for  his/her remarks that resulted in a
better presentation of this  letter. We acknowledge  financial support
of MIUR (PRIN 2001, 2002 and 2003).

 
\end{acknowledgements}

\end{document}